\documentclass[10pt,letterpaper]{article}
\usepackage{opex3}
\usepackage{amssymb,amsmath,amsfonts}
\usepackage{textcomp}
\usepackage{multirow}
\usepackage{booktabs}

\newcommand{\fref}[1]{Fig.~\ref{fig:#1}}
\newcommand{\eref}[1]{(\ref{eq:#1})}
\newcommand{\sref}[1]{$\S$\ref{sec:#1}}

\newcommand{\sub}[1]{_{\mathrm{#1}}}

\begin{document}

\title{Loss in long-storage-time optical cavities}

\author{T. Isogai, J. Miller, P. Kwee, L. Barsotti and M. Evans}

\address{LIGO Laboratory, Massachusetts Institute of Technology,  \\
  Cambridge, MA, 02139}

\email{jmiller@ligo.mit.edu} 



\begin{abstract}
  Long-storage-time Fabry-Perot cavities are a core component of many
  precision measurement experiments. Optical loss in such cavities is
  a critical parameter in determining their performance; however, it
  is very difficult to determine a priori from independent
  characterisation of the individual cavity mirrors. Here, we
  summarise three techniques for directly measuring this loss in situ
  and apply them to a high-finesse, near-concentric, 2~m
  system. Through small modifications of the cavity's length, we
  explore optical loss as a function of beam spot size over the 1-3~mm
  range. In this regime we find that optical loss is relatively
  constant at around 5~ppm per mirror and shows greater dependence on
  the positions of the beam spots on the cavity optics than on their
  size. These results have immediate consequences for the application
  of squeezed light to advanced gravitational-wave interferometers.
\end{abstract}



\section{Introduction}
Long-storage-time optical cavities are an essential tool in the quest
to expand our understanding of the universe. Currently, these devices
are employed in the world's best clocks \cite{Hinkley2013} and
frequency references \cite{Kessler2012}; experiments exploring
strong-field General Relativity, the structure of nuclear matter and
the validity of cosmological models \cite{Willke06,Harry2010,aVirgo};
searches for vacuum birefringence \cite{Valle2013} and holographic
noise \cite{Hogan2012}; and studies of macroscopic quantum mechanics
\cite{Abbott09I} and the squeezed-state rotation produced by filter
cavities \cite{Kimble2002}.

The storage time of an optical cavity is defined as the time taken for
the cavity field to decay by $1/e$. For a cavity of length
$\mathcal{L}\sub{cavity}$ and finesse $\mathcal{F}$ (see \eref{finesse}), the
storage time can be written as
\begin{equation}
\label{eq:storage}
  \tau\sub{storage}=\frac{2\mathcal{L}\sub{cavity}\mathcal{F}}{\pi c}=\frac{\mathcal{F}}{\pi f\sub{FSR}},
\end{equation}
where $f\sub{FSR}= c/(2\mathcal{L}\sub{cavity})$ is the cavity
free-spectral-range.  The storage time is therefore directly
proportional to the product of finesse and cavity length.  Optical
loss, predominantly due to scattering from the surfaces of the cavity
optics, limits the achievable finesse and storage time for a given
cavity length.

Although the surfaces of mirrors can be mapped with nm-level precision
and the absorption of bulk materials is well-known, a number of
effects prevent an accurate extrapolation from these measurements to
the optical loss which will be observed when a cavity is
formed. Dependence on spot position, coating defects, the capture of
higher-order modes and other unexplained mechanisms all conspire to
make theoretical analyses intractable. Simulations are equally
challenging, even when performed using FFT-based techniques
\cite{Vinet92}.

Previous measurements of loss in long-storage time cavities were made
on a variety of disparate systems during the course of other
investigations. These results suggest that optical loss increases with
beam spot size, implying that an almost confocal cavity is optimal
(since this geometry has the smallest beam spots for a given cavity
length). The following expression is an empirical scaling law based on
the data reported in Figure~4 of~\cite{Evans2013},
\begin{equation}
  \label{eq:fit}
  L\sub{rt}(\mathcal{L}\sub{confocal})= 10~\mathrm{ppm}\cdot \left(
    \frac{\mathcal{L}\sub{confocal}}{1~\mathrm{m}}\right)^{0.3}. 
\end{equation}
Here $L\sub{rt}$ is the round-trip loss and
$\mathcal{L}\sub{confocal}$ is the length of the confocal cavity which
has the same spot size at its mirrors as the cavity whose losses were
published (see \eref{confocal}).

\begin{figure}[thbp!]
\centering\includegraphics[width=0.7\textwidth]{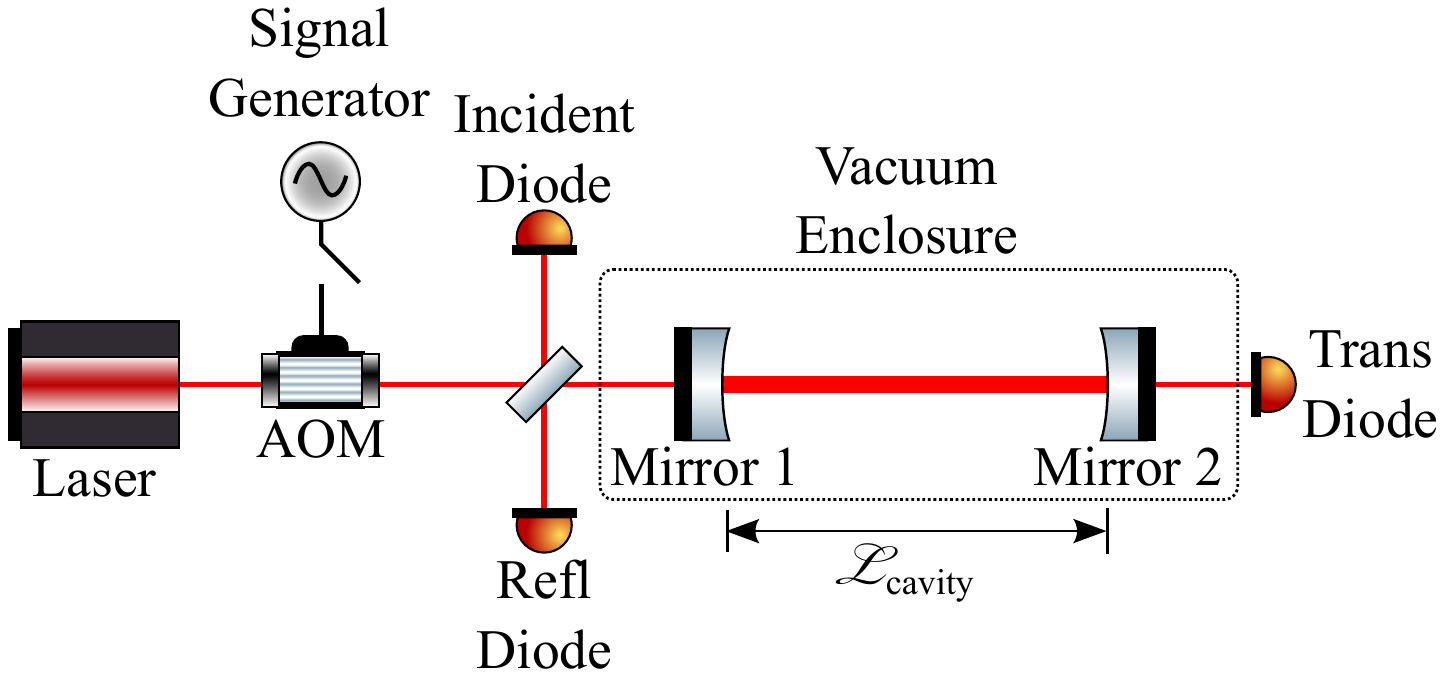}
\caption{A schematic representation of our experimental apparatus. See
  \sref{apparatus} for additional detail.}
\label{fig:layout}
\end{figure}
This approximation of the experimental results, was not based on any
physical analysis and has never been verified under controlled
conditions using coherent methods. In this work, to the best of our
knowledge, we present the first systematic study of optical loss in a
long-storage-time cavity as a function of beam spot size.

\section{Definition of cavity loss}
\label{sec:definition}
The optical properties of a mirror may be characterised by
conservation of energy,
\begin{equation}
 \label{eq:consE}
  R+T+L=R+A=1,
\end{equation}
where $R=r^2$ is the power reflectivity, $T=t^2$ is the power
transmissivity and $L$ is the coefficient of power loss. We also
introduce the attenuation, $A=T+L$, to describe the total power lost
\emph{from the cavity} upon a single reflection.

Our goal is to investigate the round-trip loss of a two-mirror cavity,
$L\sub{rt}=L_1+L_2$. Here and henceforth subscript 1 refers to the
cavity input mirror and subscript 2 to the cavity end mirror (see
\fref{layout}).

A quantity which is easily accessible via experiment is the round-trip
(amplitude) reflectivity, $r_1r_2$. Beginning from \eref{consE} and
expanding $r_1r_2$ to first order in $L$ and $T$, the round-trip loss
may be cast in the form
\begin{equation}
  \label{eq:loss}
  L\sub{rt}=2(1-r_1r_2)-(T_1+T_2).
\end{equation}
Whence we also define the round-trip attenuation
\begin{equation}
  \label{eq:attenuation}
   A\sub{rt}=A_1+A_2=2(1-r_1r_2).
\end{equation}

In practice, we are able to measure the round-trip attenuation
$A\sub{rt}$ and the input mirror transmissivity $T_1$, which, for our
strongly overcoupled cavity, are the only relevant parameters.
Therefore, we simplify our analysis by setting $r_2=1$ ($T_2=0$),
effectively incorporating the transmissivity and loss of the end
mirror into the loss of the input mirror such that
$A\sub{rt}$=$T_1$+$L_1$.

\begin{figure}[thbp!]
\centering\includegraphics[width=12cm]{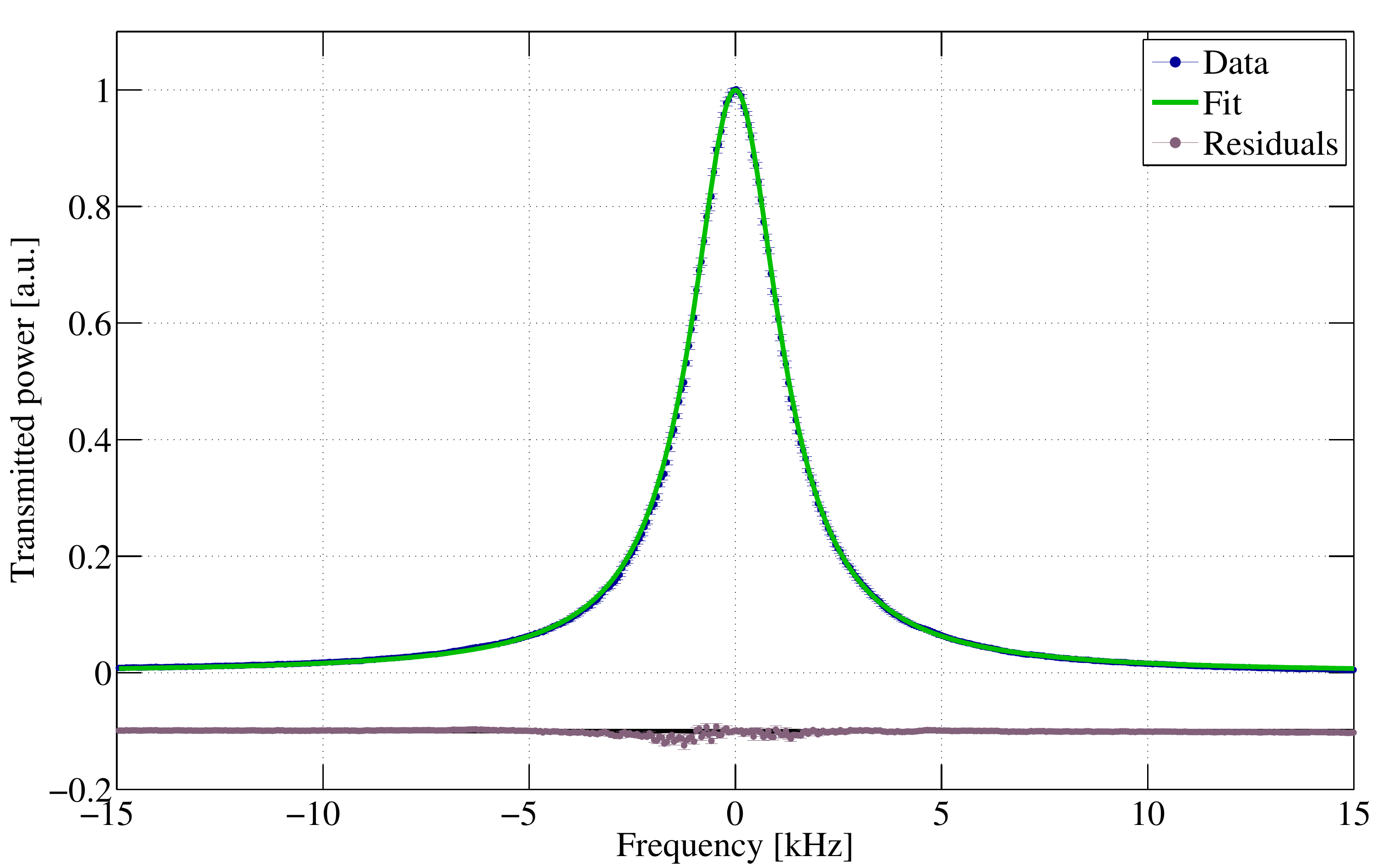}
\caption{A representative linewidth measurement. 30,000 samples were
  taken at each of 500 frequencies over a 30~kHz span. Residuals are
  offset by -0.1 for clarity.}
\label{fig:linewidth}
\end{figure} 
\section{Description of the experimental set-up}
\label{sec:apparatus}
A symmetric, near-concentric ($\mathrm{ROC}_1=\mathrm{ROC}_2\simeq
\mathcal{L}\sub{cavity}/2$) optical cavity of length
\hbox{$\mathcal{L}\sub{cavity}\lesssim2$~m} was chosen for our study. This
geometry allowed the spot size on the cavity optics to be varied over
a wide range via small ($\sim$1~cm) changes in cavity length. 

To mitigate environmental disturbances and prevent contamination, the
system was housed in a vacuum enclosure operating below $10^{-4}$~mbar
and mounted on an optical table equipped with pneumatically isolating
legs (Newport S-2000 series). The level of ambient particulates was
controlled by conducting the experiment within a class 100 soft-wall
clean room. Appropriate dress and housekeeping standards were strictly
observed.

The cavity optics consisted of ion-beam-sputtered coatings deposited
atop super-polished, two-inch, fused-silica substrates (RMS roughness
$<$1~\AA, 10-5 scratch-dig surface quality). The methods used in the
manufacture of these optics represent the current state of the
art. The measured transmissivities at our measurement wavelength were
192.2$\pm$1.3~ppm for Mirror 1 and 1.55$\pm$0.14~ppm for Mirror 2 (see
\fref{layout}); yielding a finesse of $\sim$30,000 and a linewidth of
$\sim$2.5~kHz. The optics were held in custom Siskiyou IXM200 mounts
which were actuated upon by New Focus 8301-UHV picomotors. Prior to
use, the cavity mirrors were thoroughly cleaned using spectral-grade
solvents.

The cavity input beam was generated by a 1064~nm diode-pumped
solid-state laser (JDSU M126N-1064-500). Before injection into the
cavity, the input beam was double-passed through an acousto-optic
modulator (AOM). The AOM (Gooch and Housego 3200-1113) was operated
over a 100~MHz band about its centre frequency of 200~MHz. By
modifying the frequency of the AOM drive signal, the detuning of the
input beam from resonance could be accurately controlled. Moreover, by
interrupting the AOM drive, the cavity input light could be
extinguished on a timescale of $\sim$10~ns. The power at the output of
the AOM was $\sim$1~mW.

In order to suppress cavity length noise and laser frequency
fluctuations, a frequency-shifted sample of the cavity input beam,
picked off prior to the AOM, was locked to the cavity at all times by
actuating on the laser's frequency ($\sim$40~kHz bandwidth). This
arrangement is not shown in \fref{layout}.

Readout was effected by three high-bandwidth photodetectors. Two to
sample the light transmitted and reflected by the cavity (Trans Diode
and Refl Diode respectively) and a third (Incident Diode) to
compensate for variations in laser power and the frequency dependence
of AOM output power.
%
\begin{figure}[thbp!]
\centering\includegraphics[width=12cm]{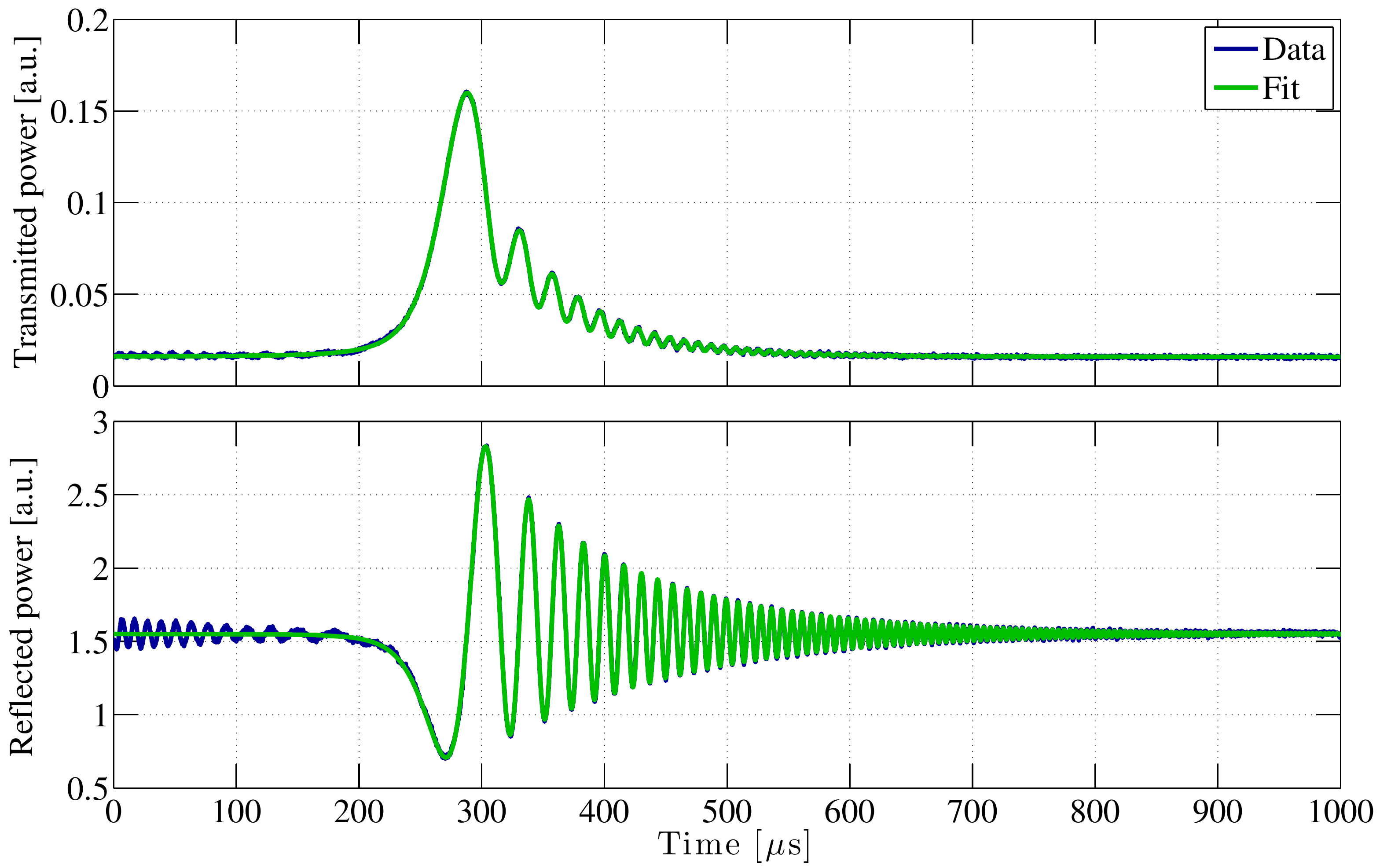}
\caption{Typical Doppler measurement. Deviations of the fit function
  (green) from recorded data (blue) at early times are explained by
  start-up transients (cavity field decaying to a steady state after
  light is injected) which are present in experiment but were not
  included in our theoretical model. This effect has no influence on
  the extracted fit parameters.}
\label{fig:doppler}
\end{figure} 
\section{Measurement techniques}
\label{sec:techniques}
We now describe the three methods used to determine the optical loss
of our cavity. Each of the techniques is able to measure round-trip
attenuation whilst the third also yields input mirror transmissivity,
allowing round-trip loss to be extracted.
\newpage
All three techniques require knowledge of the cavity
free-spectral-range, $f\sub{FSR}$. This quantity was measured to an
uncertainty of $\sim$100~Hz by noting the change in AOM drive
frequency required to move between adjacent longitudinal modes of the
cavity.

The cavity g-factor \cite{Siegman1986} was evaluated in a similar
fashion by using the AOM to map out the resonant frequencies of
higher-order spatial modes. From this description of the cavity
geometry one can calculate informative parameters such as cavity waist
size, beam spot size on the cavity mirrors and equivalent confocal
cavity length \cite{Siegman1986}. 

\subsection{Cavity linewidth}
A straightforward evaluation of the cavity
full-width-half-maximum-power linewidth, $f\sub{FWHM}$, was made by
recording the power transmitted by the cavity as the AOM slowly
stepped the input laser frequency across a fundamental-mode resonance
and fitting the result (see \fref{linewidth}). With this information
and knowledge of $f\sub{FSR}$ one can readily calculate the finesse,
\begin{equation}
\label{eq:finesse}
  \mathcal{F}=\frac{f\sub{FSR}}{f\sub{FWHM}}=\frac{\pi\sqrt{r_1r_2}}{1-r_1r_2},
\end{equation}
and therefore the round-trip attenuation (see \eref{attenuation}).

\begin{figure}[h!]
\centering\includegraphics[width=12cm]{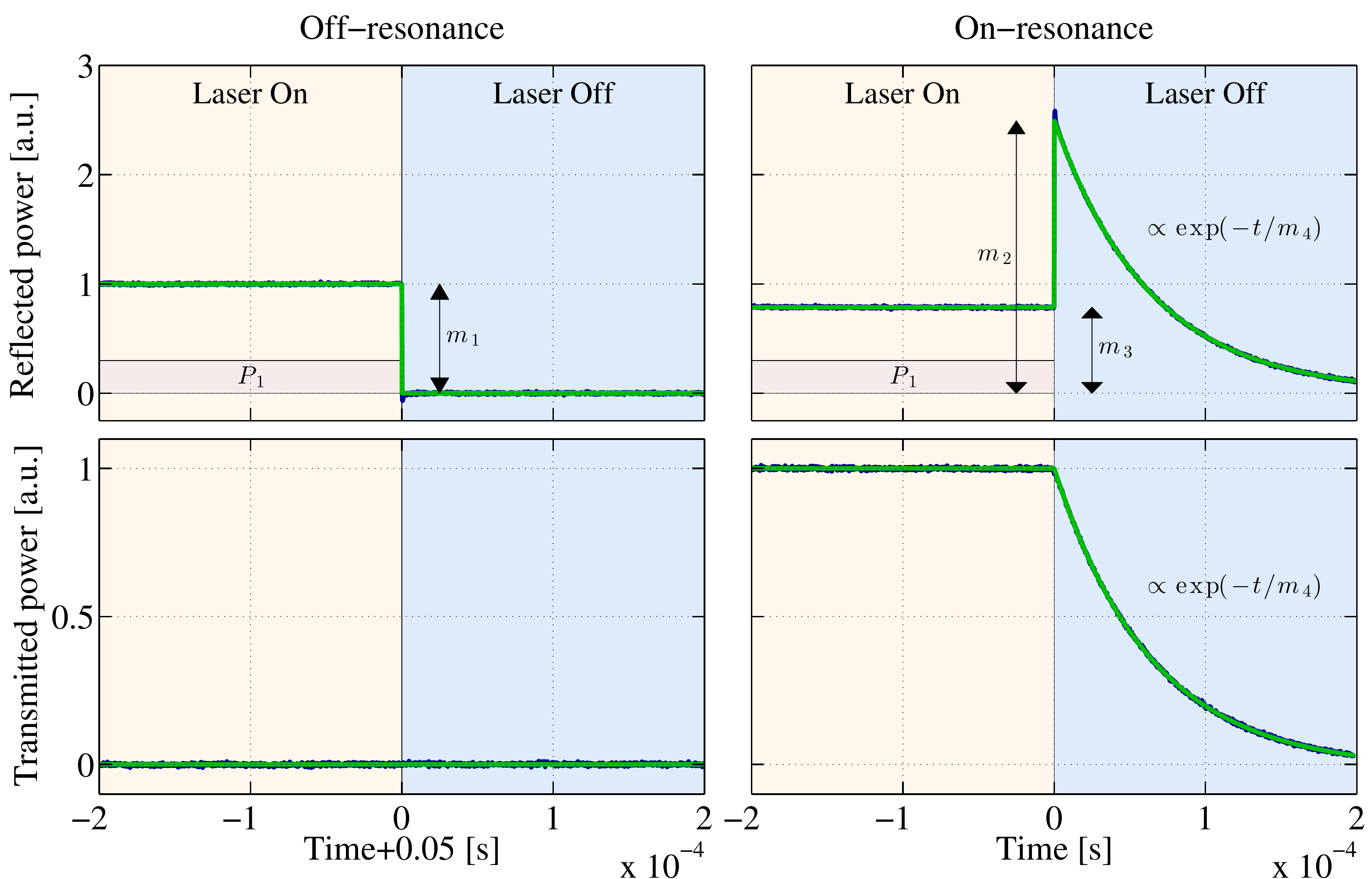}
\caption{Illustrative ringdown measurement showing reflected (upper
  axes) and transmitted (lower axes) powers. Data are shown in blue
  and theoretical fits in green. Deviations between the curves around
  $t=0$ are due to the finite detection bandwidth. A visually
  imperceptible downsampling factor of twenty five has been applied to
  the plotted data.}
\label{fig:ringdown}
\end{figure}
\subsection{Doppler}
A second, similar, method also involved tuning the laser frequency
across resonance; however, in this instance, the laser frequency was
made to sweep linearly through resonance on a timescale comparable to
the cavity storage time $\tau\sub{storage}$ ($\sim$100~\textmu s for
our system, see \eref{storage}).
\newpage
In this case, previously studied dynamical effects are observed (see
e.g.~\cite{Lawrence1999,Rakhmanov2001}), with both the transmitted and
reflected powers exhibiting damped oscillations (see
\fref{doppler}). The decay time of these oscillations is given by
$\tau\sub{storage}$, from which we again determine the cavity finesse
and round-trip attenuation.

\subsection{Ringdown}
The final technique employed capitalises on our ability to quickly
extinguish the cavity input light. Beginning from a steady-state
resonant condition, the drive to the AOM is interrupted, cutting the
input light and causing the power stored in the cavity to decay. If
one describes the cavity input power as $P\sub{in}=P_0+P_1$, where
$P_0$ ($P_1$) is that portion coupled (not coupled) into the
fundamental cavity mode, then the power circulating in the cavity $n$
round-trips after the input light is extinguished (at $t=0$) is given
by
\begin{equation*}
  P\sub{cav}(n)=P_0GT_1(R_1R_2)^n,  
\end{equation*}
where $G=g^2=[1/(1-r_1r_2)]^2$ is the cavity gain, or, extrapolating to a
function of continuous time,
\begin{equation*}
  P\sub{cav}(t\geq0)=P_0GT_1(R_1R_2)^{t/\tau\sub{rt}}=P_0GT_1\exp(-2t/\tau\sub{storage})
\end{equation*}
where $\tau\sub{rt}=1/f\sub{FSR}=2\mathcal{L}\sub{cavity}/c$ is the
cavity round-trip time and $\tau\sub{storage}$ may alternatively be
written
\begin{equation*}
  \tau\sub{storage} = -\frac{1}{f\sub{FSR}\log(r_1r_2)}\simeq \frac{g}{f\sub{FSR}}.
\end{equation*}

Therefore the transmitted and reflected powers are given by
\begin{equation*}
  P\sub{trans}(t)=
\begin{cases}
  P_0GT_1T_2 &\text{t$<$0},\\
  P_0GT_1T_2\exp(-2t/\tau\sub{storage})
  & \text{t$\geq$0}
\end{cases}
\end{equation*}
and
\begin{equation*}
  P\sub{refl}(t)=\begin{cases}
    P_0G[r_1-r_2(T_1+R_1)]^2+P_1& \text{t$<$0},\\
    P_0GT_1^2R_2 \exp(-2t/\tau\sub{storage}) & \text{t$\geq$0},
\end{cases}
\end{equation*}
where, in both cases, the expressions for $t<0$ are those of the
standard steady state \cite{Siegman1986}.



In both transmission and reflection one can easily fit the decaying
portion of these waveforms to extract $\tau\sub{storage}$ and
therefore quantify the round-trip attenuation -- a traditional
ringdown measurement. However, in reflection additional information is
available which makes it possible to also extract input mirror
transmissivity.

Experimentally, two synchronised square-wave modulations were applied
to the AOM drive such that the laser input light was 1) switched on
and off at a frequency $f\sub{mod}=40$~Hz and 2) made to alternate
between resonant and off-resonance tunings at a frequency of
$f\sub{mod}/2$. In combination, these modulations allowed us to record
the diode outputs as the cavity transitioned between four states:
off-resonance laser-on, off-resonance laser-off, on-resonance laser-on
and on-resonance laser-off. From these data we
extracted photodetector offsets and the four quantities indicated in
\fref{ringdown},
\begin{align*}
  m_1&=P_0+P_1,\\
  m_2&=P_0GT_1^2R_2,\\
  m_3&=P_0G[r_1-r_2(T_1+R_1)]^2+P_1,\\
  m_4&=\tau\sub{storage}/2.
\end{align*}

These measurements define four equations in five variables ($P_0$,
$P_1$, $r_1$, $t_1$ and $r_2$). In order to compute a solution we
again choose to set $r_2=1$ (see \sref{definition})\footnote{For
  cavities with matched coatings one can take
  $r_1=r_2$. Alternatively, performing a second set of measurements
  after swapping Mirror 1 and Mirror 2 allows one to immediately solve
  the system.}.

From this solution we can construct not only the round-trip
attenuation, as before, but also the round-trip loss,
$L\sub{rt}=A\sub{rt}-t_1^2$, and the input mode-matching fraction,
$P_0/(P_0+P_1)$. Moreover, the $t_1$ value obtained in this way can
also be used to convert round-trip attenuation values obtained via
other measurement techniques into round-trip loss.

On average, this method of obtaining $t_1$ was found to give results
consistent with independent transmissivity measurements
\cite{Zhang2011}. However, the ringdown-based technique is preferred
as it probes the same region of each mirror's surface as the
attenuation measurement, accounting for any spatial variation in
transmissivity even as the cavity axis moves. In contrast, the
two-dimensional transmissivity map of a witness sample, or even an
installed optic, can produce only an informed estimate for use in
analysis.

\section{Results}
The length of our cavity was varied from 1.932~m to 1.998~m in four
discrete steps. This range of lengths provided beam spot sizes
extending from 1.34~mm to 2.74~mm. Examination of larger spot sizes
was prohibited by cavity instability.

At every cavity length, several different alignments were
investigated. Each alignment resulted in a new cavity axis, displaced
by approximately one beam spot size at both mirrors. Explorations of
larger axis shifts produced no appreciable difference in recorded
results.

This combination of length and alignment changes allowed us to
investigate optical loss as a function of beam spot size in a
systematic way, accounting for the possible influence of coating
inhomogeneities and defects.

For every configuration, the techniques detailed in \sref{techniques}
were each applied one hundred times, mitigating the effects of random
noise and enabling drifts to be investigated and eliminated.

\subsection{Attenuation measured by each technique}
\begin{figure}[thbp!]
\centering\includegraphics[width=12cm]{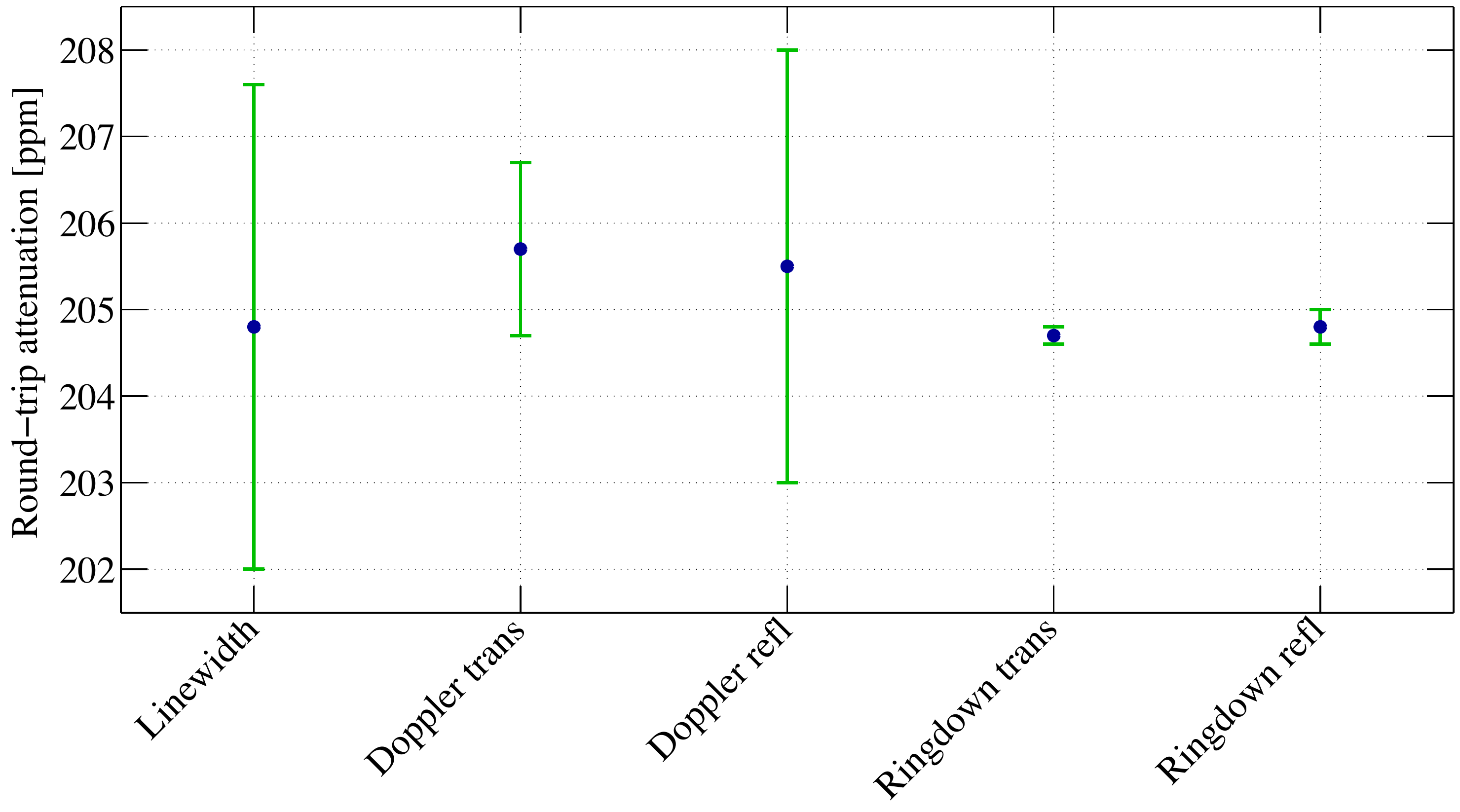}
\caption{Round-trip attenuation measured using five techniques at a
  single cavity length and alignment. Uncertainties represent the
  statistical error of one hundred measurements.}
\label{fig:lossVtechnique}
\end{figure}

Typical results for a single measurement configuration are shown in
\fref{lossVtechnique}. The mean round-trip attenuation values from all
techniques agree to better than 1~ppm. This uniformity was preserved
at all alignments and cavity lengths, engendering confidence in our
experimental method and analysis pipelines.

Achieving such consistency demanded the painstaking examination and
elimination of experimental noise sources (particularly troublesome
were scattered light and indiscernible beam clipping) and systematic
effects (cavity-induced polarisation rotation \cite{Jacob1995},
electronic offsets and pressure and temperature changes in the vacuum
vessel).

The relative magnitudes of the uncertainties may be explained by the
timescale over which the experimental data were captured. Linewidth
measurements were made comparatively slowly and were therefore more
susceptible to any unintended detunings from resonance. In our set-up,
such detunings were mainly due to residual seismic and acoustic
disturbances. Doppler and ringdown measurements were performed much
more quickly and were hence less sensitive to low-frequency
environmental noise, resulting in smaller variations within a
measurement set.

In addition to being performed most quickly, the small uncertainties
offered by ringdown measurements may further be ascribed to the
non-resonant nature of the technique -- after input light has been
extinguished, cavity length fluctuations are no longer able to
influence ringdown results.

\subsection{Attenuation and transmissivity as function of beam spot
  position}
\begin{figure}[thpb]
\centering\includegraphics[width=12cm]{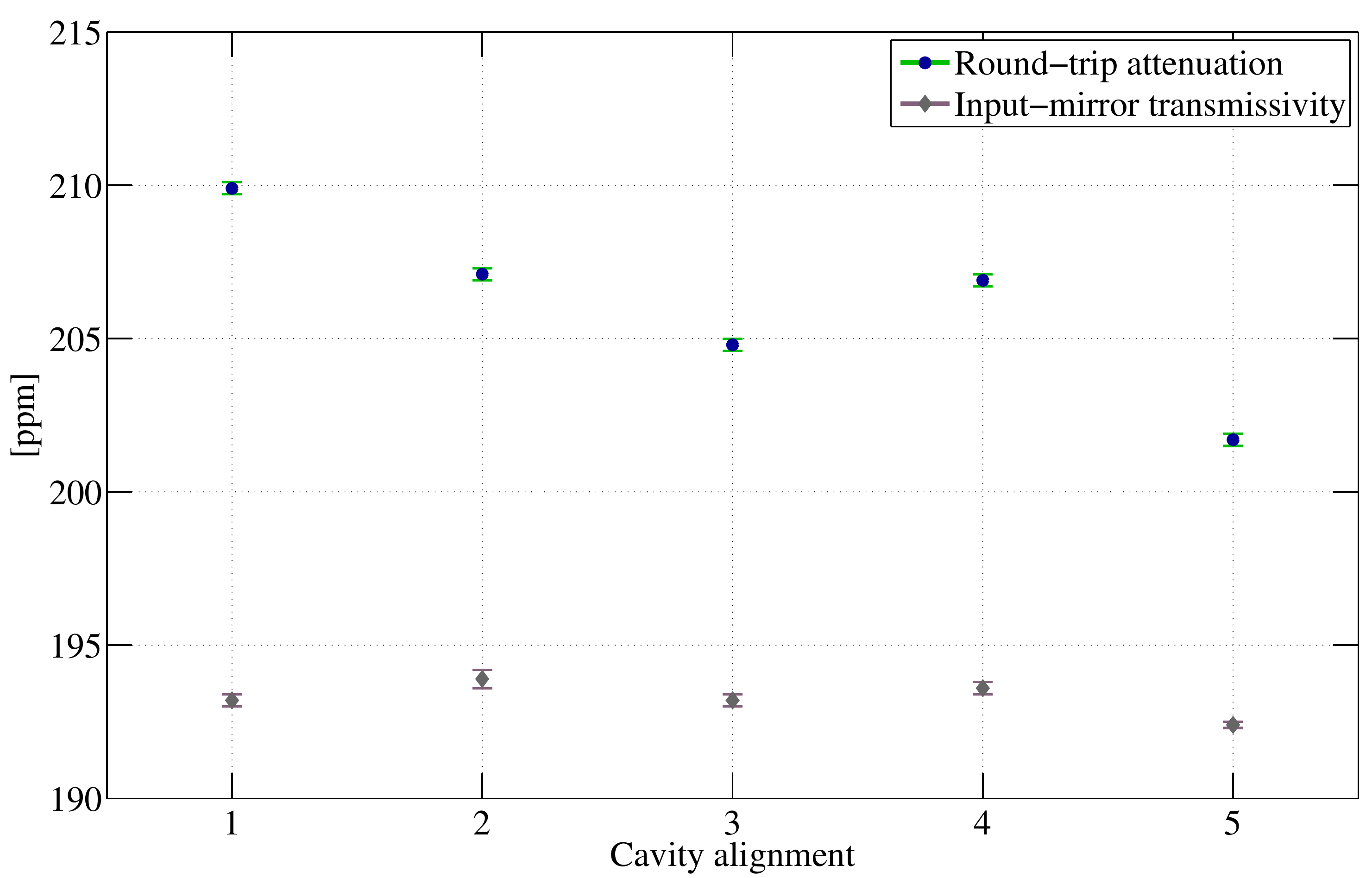}
\caption{Round-trip attenuation and input-mirror transmissivity
  obtained via ringdown measurements in reflection at five different
  cavity alignments.  Uncertainties represent the statistical error of
  one hundred measurements.}
\label{fig:lossVspotPos}
\end{figure}
In \fref{lossVspotPos} we show how the outcome of one measurement
technique (ringdown in reflection) varied as a function of alignment
at a single cavity length. We observe that the round-trip attenuation
undergoes changes more than one order of magnitude larger than its
experimental uncertainty whilst the simultaneously measured input
mirror transmissivity remains relatively constant. Equivalent
behaviour was witnessed in all of our data. Thus, beam spot position
had a strong influence on round-trip optical loss. Independent
measurements of scatter and absorption as a function of beam position
were not available to correlate against our measured data.

\begin{figure}[thpb]
\centering\includegraphics[width=12cm]{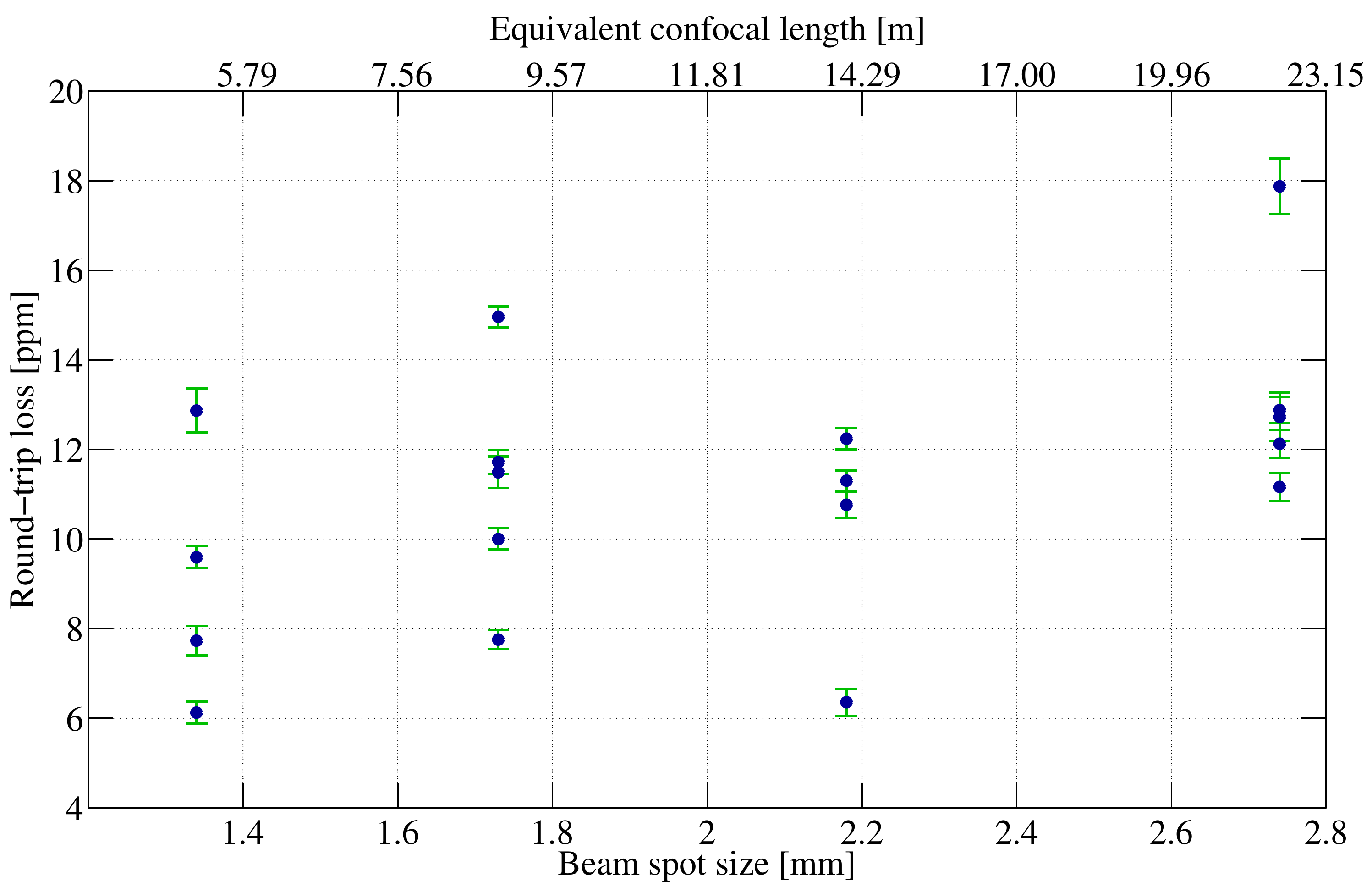}
\caption{Measured round-trip loss as a function of beam spot size and
  equivalent confocal length. Each data point corresponds to the
  combined outcome (weighted by statistical uncertainty) of three
  different measurement techniques applied at a single cavity length
  and alignment. The resulting experimental error is almost equally
  distributed between uncertainty in determination of $T_1$, $T_2$ and
  $A\sub{rt}$.  Several alignments were explored at each cavity
  length, yielding multiple loss values for each beam spot
  size. Equivalent confocal lengths are rounded to two decimal
  places.}
\label{fig:loss}
\end{figure}

\begin{figure}[thpb!]
\centering\includegraphics[width=12cm]{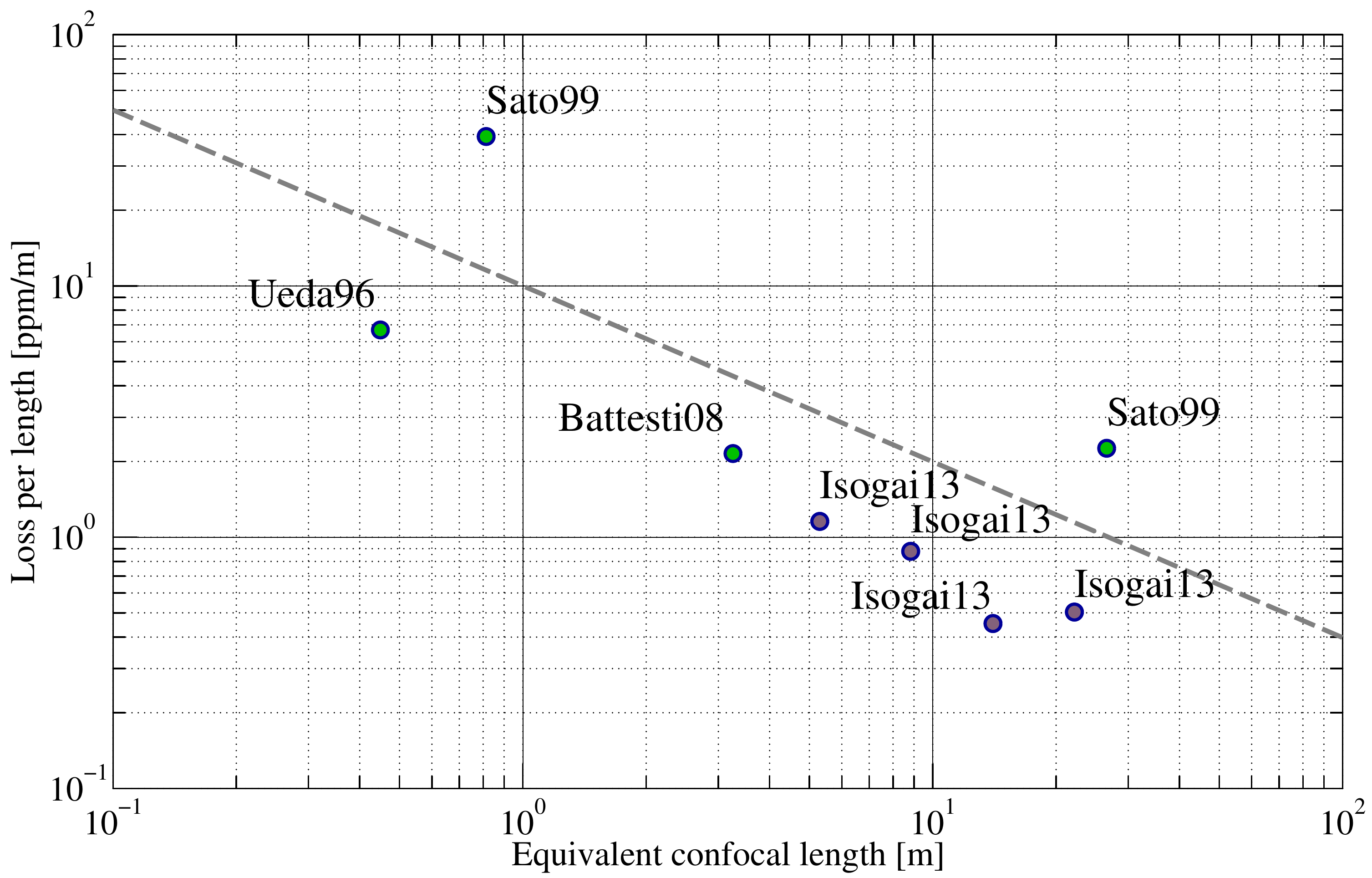}
\caption{Loss per unit length as a function of equivalent confocal
  length. Our results (Isogai13, blue and violet markers) are compared
  to previously published works. References are available in Figure 4
  of \cite{Evans2013}, from where this plot was reproduced with
  permission.}
\label{fig:literature}
\end{figure}
\subsection{Round-trip loss as function of beam spot size and
  equivalent confocal length}
Results from all measurement techniques, cavity lengths and spot
positions are combined in \fref{loss} to show round-trip optical loss
as a function of beam spot size.

In addition to utilising all information available from experiment, we
also subtract the value of $T_2$ (1.55$\pm$0.14~ppm), obtained via
independent examination of a coating witness sample \cite{Zhang2011},
from our estimate of $L\sub{rt}$. $T_2$ cannot be directly measured in
our set-up but tacitly contributes to the round-trip attenuation
$A\sub{rt}$ as we chose to set $r_2=1$.
 
In \fref{loss} we also characterise our results in terms of the
equivalent confocal cavity length
\begin{equation}
\label{eq:confocal}
  \mathcal{L}\sub{confocal}=\frac{\pi w^2}{\lambda\sub{laser}},
\end{equation}
where $w$ is the beam spot size on the cavity optics and
$\lambda\sub{laser}$ is the wavelength of the input laser light. For a
given spot size, a confocal geometry gives the longest possible cavity
and therefore the narrowest linewidth, the longest storage time and
the best frequency or length discrimination. Numerical simulations of
ten concentric cavities and their confocal equivalents, with spot
sizes ranging from 1.2~mm to 3.9~mm and realistically imperfect
optics, revealed no significant geometry-dependent variations in loss.

It should be noted that, due to alignment sensitivity, data taken when
the beam spot size was at its largest (2.74~mm) were not of a quality
equivalent to the remainder of the data. Whilst the values presented
are an accurate representation of the measured loss when the cavity
alignment was optimal, small perturbations resulted in values larger
by tens of ppm, presumably due to beam clipping. For the same reason,
linewidth measurements proved unreliable for the largest spot size and
were not incorporated into our analysis.
\newpage
To set our results in context, we plot our lowest-loss results for
each cavity length together with values obtained from the literature
(see \fref{literature}). Data are described in terms of loss per unit
length, where, for the reasons discussed above, the length is that of
the equivalent confocal cavity.

\section{Discussion and conclusions}
\label{sec:discussion}
In this work we have performed the first, to our knowledge, systematic
investigation of optical loss as a function of beam spot size. We find
that our cavity mirrors, representative of the best available at this
time, offer an approximately constant round-trip optical loss of
$\sim$10~ppm (5~ppm per bounce) for beam spot sizes in the 1-3~mm
range (equivalently, confocal lengths in the 5-25~m range). It is
important to note, however, that a significant range of optical losses
($\sim$6~ppm to $\sim$18~ppm) was observed as the position of the beam
spots on the cavity mirrors was varied.

From our measurements we conclude that variations in loss result from
scattering caused by point defects with an average separation larger
than a few millimetres. This conclusion is supported by direct
measurements of the scattering produced by our cavity mirrors, made
following the methods described in \cite{Sandoval2012}. The total
integrated scatter of each of our optics (measured using a 4-6~mm beam
spot) was estimated to be $\sim$10~ppm \cite{Smith2013}, comparable to
the per-bounce loss inferred from the data presented in \fref{loss}.

The results presented here are of immediate interest in relation to
ground-based interferometric gravitational-wave detectors
(e.g.~Advanced LIGO \cite{Harry2010}). Squeezed states of light
\cite{Gerry2005} have recently been shown to enhance the sensitivity
of gravitational-wave interferometers by reducing quantum shot noise
at high frequencies ($\gtrsim$100~Hz)
\cite{Schnabel2011,Barsotti2013}. Future interferometers require that
the squeezed quadrature be rotated as a function of frequency.  Such
rotation can be realised by reflecting the squeezed light from
\emph{filter cavities} -- long-storage-time optical cavities, operated
in a detuned configuration, whose linewidth is comparable to the
frequency range over which the squeezed quadrature must be rotated
($\sim$50~Hz) \cite{Kimble2002}. A recent investigation, using
realistic parameters and noise estimates, concluded that a two-mirror
cavity with an optical loss of 1~ppm/m would be sufficient to realise
an effective filter cavity for an Advanced-LIGO-like interferometer
\cite{Evans2013}.

Our measurements show that a confocal cavity approximately 10-20~m in
length fulfils this requirement. Such a cavity can be constructed
within the existing vacuum envelope of the LIGO interferometers
without resorting to use of the long arm cavities. Frequency dependent
squeezing is thus an attractive and viable near-term upgrade for
Advanced LIGO.

Optical loss in long-storage-time cavities is a key parameter across a
wide variety of disciplines. Our results demonstrate that, using
currently available optics, 10 m-scale cavities with storage times of
order 10~ms are now within reach.

\section*{Acknowledgements}
The authors gratefully acknowledge the support of the National Science
Foundation and the LIGO Laboratory, operating under cooperative
Agreement No. PHY-0757058. They also acknowledge the guidance and
invaluable input of Peter Fritschel and Nergis Mavalvala and fruitful
discussion and collaboration with Josh Smith, Valery Frolov, Hiroaki
Yamamoto and Jan Harms. Mirror transmissivity measurements were made
possible through the assistance and expertise of Eric Gustafson and
Liyuan Zhang. Myron MacInnis assisted in construction and maintenance
of the experimental apparatus. This paper has been assigned LIGO
Document No. LIGO-P1300159.

\end{document}